# Multiple Power Quality Event Detection and Classification using Modified S Transform and WOA tuned SVM Classifier

Sambit Dash* , Umamani Subudhi

Department of Electrical Engineering, IIIT Bhubaneswar, Gothapatna, Odisha, India.

Corresponding Author-*a417001@iiit-bh.ac.in

**Abstract-**In this paper, a novel method for classification of power quality events is illustrated. 15 types of power quality events consisting of single and multi-stage disturbances are considered for study. A database of the synthetic PQ events is generated in MATLAB using mathematical models. The generated signals are passed through a novel Modified Stockwell transform consisting of second order gaussian window which provides the ST matrix. From the ST matrix various statistical features such as energy and standard deviation of the magnitude and phase contour are extracted and given as input to Support Vector Machine (SVM). Furthermore, to improve the performance of SVM, a novel meta-heuristic technique called Whale Optimization Algorithm (WOA) is used to tune the hyper parameters of the SVM classifier. The performance of the proposed method is analyzed under noisy and noiseless conditions. It is observed that WOA tuned SVM provides improved classification accuracy than other widely used meta-heuristic optimization algorithms such as Particle Swarm Optimization (PSO) tuned SVM and Genetic Algorithm (GA) tuned SVM. Further, two novel circuits for generation of sag, swell and interrupt are developed and the proposed technique is validated on real time signals obtained from the circuits.

## 1.Introduction

In recent years power quality has become a major concern in electric grids. Improper or bad power quality can cause catastrophic financial losses in semiconductor and paper processing industries. There has been an exponential growth of renewable energy sources and increased use of nonlinear electronic loads. Power converters used in such energy sources and electronic loads tend to inject current and voltage harmonics into the electric grid which can have major repercussion on sensitive devices connected to the grid elsewhere. In this regard proper detection, classification and mitigation of such power quality events is of utmost importance. Advancement in various digital signal processing techniques have led to deployment of such techniques for detection of power quality events. Similarly, growth in the field of machine learning have helped in automatic classification of such events. Generally, a signal is fed to a digital signal processing technique for feature extraction. The extracted features are then used to train a machine learning algorithm which is used to classify different power quality events. The accuracy of such techniques depends purely on the signal processing technique adopted, the extracted features and the specific classifier chosen. Since neither any single approach for detection and classification can be considered as best, thus the problem of detection and classification continues to be an area of active research.

In literature, various signal processing techniques are illustrated such as Fourier transform (FT) [1], Short time Fourier transform(STFT) [2] , Wavelet transform (WT) [3], Stockwell's transform (S-transform) etc. When a signal is passed through FT it gives information regarding the frequency spectra in the signal without providing time information at which such frequencies occur. STFT tries to overcome the limitation of FT by dividing the signal into various independent windows and then FT is applied to find the frequency content of each window. Although STFT provides time localization but the accuracy of such localization depends on the window width selected. To overcome the disadvantages of STFT, a multi resolution technique called Wavelet transform (WT) is adopted. WT acts as a filter decomposing signals into high frequency and low frequency components. The approximate coefficient refers to the low frequency component and the detail coefficient refers to the high frequency component. Although WT provides better time and frequency localization yet its performance depends on the presence of noise in the signal and choice of mother wavelet. S transform is a modification of continuous WT. It provides a frequency dependent resolution of decrease in width with increase in frequency. S transform is fairly robust to noise in the signal which makes it ideal for the detection of PQ in real time signals. For classification, various statistical machine learning techniques can be applied.

Mishra et al. [4] used a combination of S- transform and Probabilistic Neural Network (PNN) for classification of PQ events. A combination of decision tree and artificial neural network was utilized by Kumar et al. [5] for classification of power quality events. Reddy et al. [6] used an ensemble of decision trees called random forest for classification and compared its performance with different algorithms. Rajeshbabu et al. [7] used linear discriminant analysis for feature extraction and an analytic hierarchy method for classification. Wang et al. [8] have utilized deep convolutional neural network to automatically extract features and classify power quality events but such technique requires massive computing resources. Shaik et al. [9] uses S transform for detection of power quality events in renewable energy integrated power system and framed a power quality index of various events is generated which is used to determine the type of PQ event.

In this paper Whale Optimization Algorithm tuned Support Vector Machine is used for classification of signals. SVM is a structural risk minimization algorithm utilizing a definite hyper plane for multi-class classification. Although SVM is capable of high accuracy classification in power quality domain, but the choice of hyper parameters is crucial for such models. Manual selection of hyper parameters may not provide optimal solution especially in case of radial basis function (RBF) kernel which have more than one hyper parameters to tune. To overcome this problem a novel meta-heuristic optimization technique called whale optimization algorithm (WOA) is utilized for hyper parameter selection of SVM. WOA is based on hunting pattern of the humpback whales. It has been illustrated that WOA has superior accuracy and performance capability in various optimization problems [10]. This makes WOA suitable for tuning of hyper parameters of SVM.

In section 2 of this paper, a brief introduction to Stockwell's Transform is given along with the mathematical derivation and correlation of Stockwell transform to wavelet transform. Further, a mathematical background to a second order Gaussian Window S-Transform and its difference from generalized S transform is illustrated. Support Vector Machine (SVM) and its structure is described in section 3. The origin of the whale optimization algorithm and its mathematical formulation is described in section 4. Flowchart for detection, classification and SVM hyper-parameter optimization using meta-heuristic algorithm is discussed in section 5. The different types of power quality signals that are used in this paper along with their mathematical equation is also described in section 5. The S-transform contour of different Power Quality events and the classification accuracy of various classifiers is discussed in section 6.1. In section 6.2, two new circuits for generation of sag, swell and interrupt are discussed and their MATLAB simulation are illustrated. In section 6.3, laboratory setup of sag, swell and interrupt generator circuits are shows with real time signal collection and application of proposed technique.

## 2. Second Order Gaussian Window Stockwell's Transform (SOGW-ST)

In recent years S-transform [11,12] has evolved as a prime signal processing technique for non-stationary signals. S-transform can be considered as a modification of continuous wavelet transform (CWT) as it also analyzes data in the phase component of CWT by modifying the phase aspect of the mother wavelet. It acts as a set of filters conducting multi resolution analysis with constant relative bandwidth. The unique feature of S-transform is the addition of frequency dependent resolution while simultaneous localization of real and imaginary spectra. Analyzing the phase and amplitude spectra of this transform provides both time and frequency localization.

*Continuous S-Transform:* For a function h(t), the CWT W($\tau, d$) can be defined by the equation 1.1

$$w(\tau, d) = \int_{-\infty}^{\infty} h(t) w(t - \tau, d) dt \quad \ldots\ldots(1.1)$$

The scale parameter which controls resolution is denoted $d$ and the $\tau$ gives the position of wavelet. The S-transform is basically CWT of distinct mother wavelet which is multiplied by a phase factor denoted in equation 1.2

$$S(\tau, d) = e^{-i2\pi f \tau}(t) w(\tau, d) \quad \ldots\ldots..(1.2)$$

where for the given signal the mother wavelet $w(\tau, d)$ can be denoted by equation 1.3

$$w(\tau, d) = \frac{f}{\sqrt{2\pi}} e^{-\frac{t^2 f^2}{2}} e^{i2\pi ft} \quad \ldots\ldots\ldots(1.3)$$

The Gaussian window width is represented as $\sigma(f) = T = 1/|f|$

The FT representation of S-transform is denoted by the equation 1.4

$$S(\tau, d) = \int_{-\infty}^{\infty} H(\alpha + f) e^{-\frac{2\pi^2 \alpha^2}{f^2}} e^{i2\pi ft} \quad \ldots\ldots(1.4)$$

*Second Order Gaussian Window:*
It must be noted that a Gaussian window whose standard deviation changes proportionally tends to be problematic for analysis of many real time signals of high frequency and large duration also low frequency of small duration hence various types of Gaussian windows were proposed by [13] [14] [15]. In this paper we propose a Gaussian window which is a second order function of frequency denoted by eq. 1.5

$$\sigma(f) = \frac{af^2 + bf + c}{f} \quad \ldots\ldots\ldots(1.5)$$

For the purpose of detection of power quality events where the frequency range of 0 to 1KHz, the value of 'a' is 6, 'b' is 12 and 'c' is 0.08. The choice of a, b and c must be on the type of application and the range of frequencies present in the signal.

In the given case the second order gaussian window can be given as in eq. 1.6:

$$w(\tau - t, f) = \frac{f}{(af^2 + bf + c)\sqrt{2\pi}} e^{\frac{-(t-\tau)^2 f^2}{2(af^2 + bf + c)^2}} \quad \ldots\ldots\ldots(1.6)$$

Using the proposed window the modified S-Transform (MST) called as Second Order Gaussian Window S-Transform (SOGW-ST) can be given as :

$$S_x^{a,b,c}(\tau, f) = \int_{-\infty}^{\infty} x(t) \frac{|f|}{(af^2 + bf + c)\sqrt{2\pi}} e^{\frac{-(t-\tau)^2 f^2}{2(af^2 + bf + c)^2}} e^{-i2\pi ft} dt \quad \ldots\ldots(1.7)$$

The proposed modified S transform satisfies the normalization and invertibility property.

### 3. Machine learning Classifier

### i) Support Vector Machine:

Support Vector Machine (SVM) is a hyper plane based classifier adopting a structural risk minimization approach [16,17]. SVM consists of hyper plane created using various data points of different classes of data and these data points are called as support vectors. The superiority of SVM compared to other machine learning algorithm is the ability to generalize data when transformed into a high dimensional feature space.

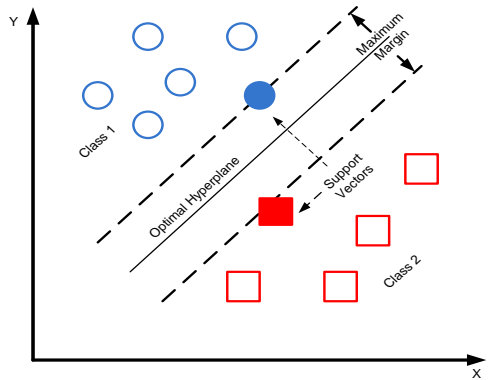

Figure 1: General Structure of SVM classifier

The hyper plane of SVM can be illustrated by (2.1)

$$f(s) = \mathbf{w}^T s + b = \sum_{j=1}^{n} w_j s_j + b \quad \ldots\ldots(2.1)$$

where w can be considered as the weight vector of the first order linear equation and b is the constrained bias of the linear equation .

The width of the hyper plane can be changed by changing the values of w, b and the optimal hyper plane separating different classes of data can be defined as the following optimization problem:

$$\text{minimize } \tfrac{1}{2}\|w\|^2 + C\sum_{i=1}^{M}\xi_i \qquad \ldots\ldots(2.2)$$

subject to

$$o_i(w^T s + b) \geq 1 - \xi_i, \text{ for i = 1,2,....M} \qquad \ldots\ldots(2.3)$$

The equation defining optimal value of b is

$$b^* = -\tfrac{1}{2}\sum_{SVs} o_i \alpha_i^* (\upsilon_1^T s_i + \upsilon_2^T s_i + ..) \qquad \ldots(2.4)$$

where $\upsilon$ denote the support vectors

The decision function can be illustrated as

$$f(s) = \sum_{SVs} \alpha_i o_i s_i^T s + b^* \qquad \ldots(2.5)$$

In this paper, RBF is used as the activation function. The hyper parameters of SVM classifier for RBF activation function are 'c' and 'gamma'. Initially, for manual tuning of SVM, 'c' is varied from 1 to 9000 and 'gamma' is varied from 1 to 1000. Then for an optimal solution, WOA algorithm is utilized to find the optimal values of 'c' and 'gamma'.

## 4. Whale Optimization Algorithm (WOA)

WOA is a novel meta-heuristic nature-inspired approach for optimization developed by Mirjalili [10]. WOA simulates an interesting pattern observed in the social behavior of humpback whales. Although humpback whales are some of the largest mammals in the planet it has been observed that these animals have unique intelligence level and capability of sophisticated collective work. Typically these animals stay in small groups and prey on krill and small fish herds. A unique behavior of bubble net feeding technique is observed in such animals. It is a complexly coordinated tactic of catching multiple preys at once by initially emitting high pitch calls which panic the prey to move towards the surface. Then the humpback whales in a circular coordinated manner corner the prey and finally attack it.

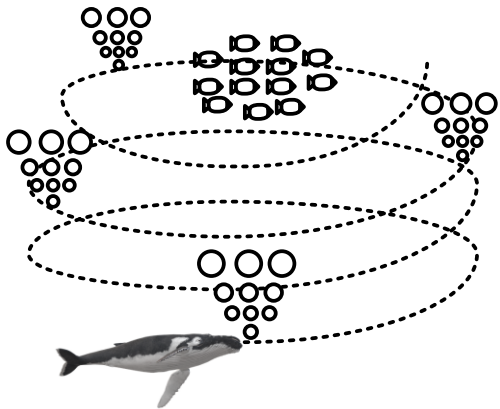

Figure 2: Bubble net feeding technique of humpback whales

Mathematically, WOA can formulated as follows

*i) Shrinking encircling technique:*

While hunting, initially the whales encircle the target prey which is considered as the best candidate solution. Once the best search agent is decided other search agents update their position accordingly. This can be represented by the following equation:

$$\vec{D} = |\vec{C}.\vec{X}^*(t) - \vec{X}(t)| \quad \ldots(3.1)$$

$$\vec{X}(t+1) = (\vec{X}^*(t) - \vec{A}.\vec{D}) \quad \ldots(3.2)$$

where t indicates the current iteration and A, C are the coefficient vectors and $X^*$ is the position vector of best solution and X is the current position vector. A and C can defined as follows:

$$\vec{A} = 2\vec{a}.\vec{r} - \vec{a} \quad \ldots(3.3)$$

$$\vec{C} = 2\vec{r} \quad \ldots(3.4)$$

where $\vec{a}$ is linearly decreased from 2 to 0 and $\vec{r}$ is a random vector between 1 and 0.

*ii) Spiral updating position*

The equation illustrating the spiral position update between the humpback whale and the prey is given as

$$\vec{D'} = |\vec{X}^*(t) - \vec{X}(t)| \quad \ldots(3.5)$$

$$\vec{X}(t+1) = \vec{D'}.e^{bl}.\cos(2\pi l) + \vec{X}^*(t) \quad \ldots(3.6)$$

where b defines the shape of the spiral and $l$ is a random number between [-1, 1]

To find the global optima, the position of search agent is updated randomly. The mathematical model can be defined as

$$\vec{D} = |\vec{C}.\vec{X}^*(t) - \vec{X}(t)| \quad \ldots(3.7)$$

$$\vec{X}(t+1) = (\vec{X}_{rand} - \vec{A}.\vec{D}) \quad \ldots(3.8)$$

where $X_{rand}$ is chosen randomly from current trial.

For SVM classifier with Radial Basis Function (RBF) based activation function there are 2 hyper parameters to tune namely 'c' and 'gamma'. In each of the meta-heuristic algorithm the search agents are utilized to find the optimal value of hyper parameters which provide the maximum accuracy within the search space. The cost function is accuracy of validation dataset which is to be maximized. In each iteration, the values of 'c' and 'gamma' are updated and the corresponding accuracy is found. The upper and lower bound of both the parameters are fixed between 200000 and 0.01. The maximum number of iterations chosen is 100 and the number of search agents chosen is 10.

## 5. Proposed Methodology

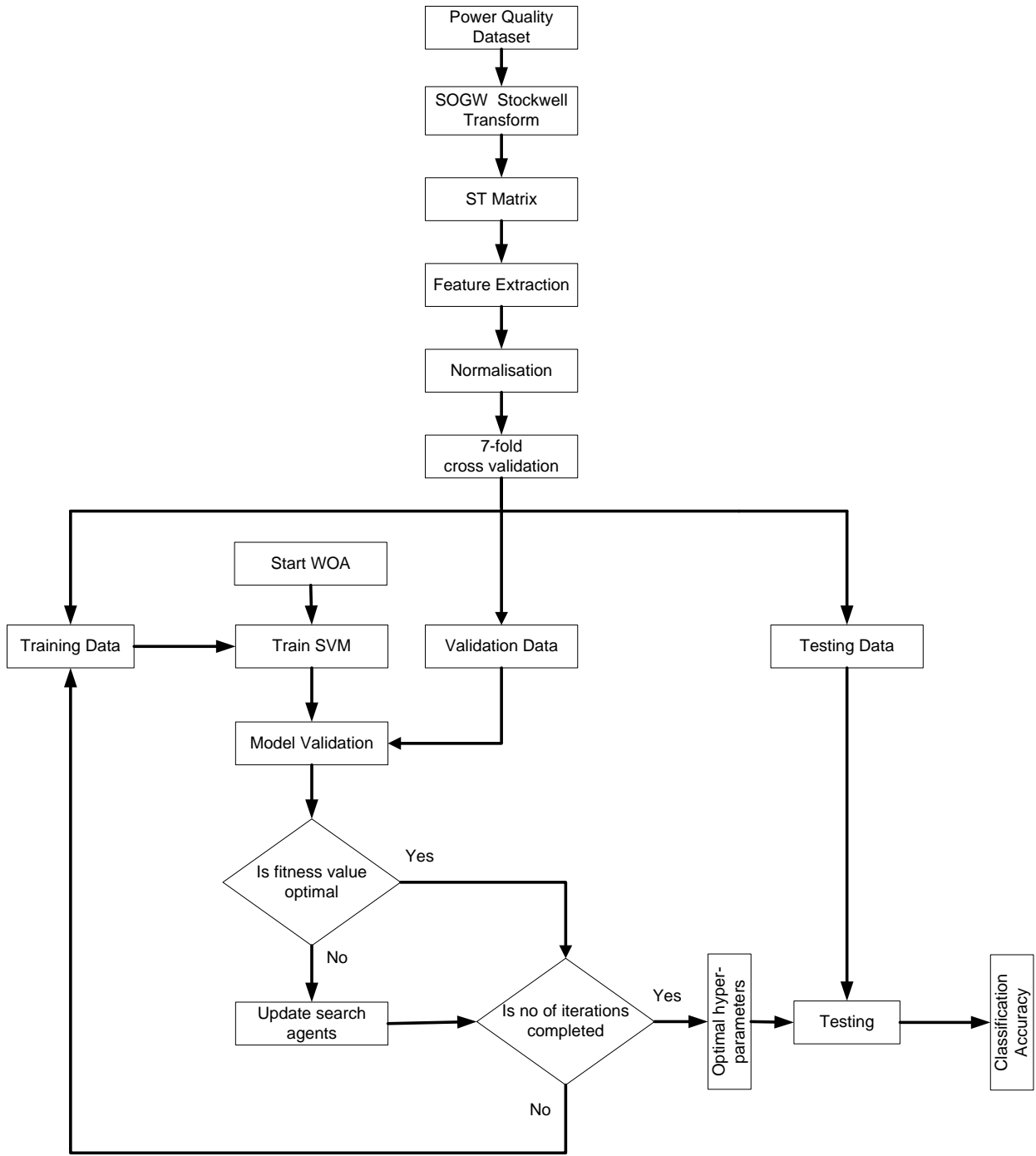

Figure 3 : Flow Chart of Power Quality Detection and Hyper Parameter Optimization of SVM Classifier

Figure 3 illustrates a step wise flow chart from the point of power quality signal generation till obtaining the classification accuracy of tuned SVM classifier. To verify the proposed approach, single and multi-stage PQ disturbance signals are generated using MATLAB 2014A. For proper training of classifier, it is advantageous to use synthetic data as multiple variations of the same disturbance can be easily recreated artificially. Mathematical model of different PQ events used in this paper [18,19,20] are given in table 1. The mathematical formulas are coded in MATLAB and a for loop is run for 100 iterations with change in the various parameters of each type of PQ event thereby generating a comprehensive database of PQ signals. A sampling frequency of 3.2 KHz is taken in this paper. Signals consisting of more than 2 disturbances are included in the dataset. The generated PQ signals are passed through SOGW Stockwell Transform (SOGW-ST) which in turn generates ST Matrix. The amplitude and phase information are contained in the ST matrix. From the ST matrix, various features are extracted. The standard deviation of magnitude vector is taken as feature 1, the energy of magnitude vector as feature 2, the standard deviation of frequency vector as feature 3, the energy of frequency as feature 4. Feature set is 7 fold cross validated. Then the dataset is divided in the ratio of 70:10:20, where 70 signals of each type of disturbance are used for training, 10 are used for validation and 20 signals are used for testing.

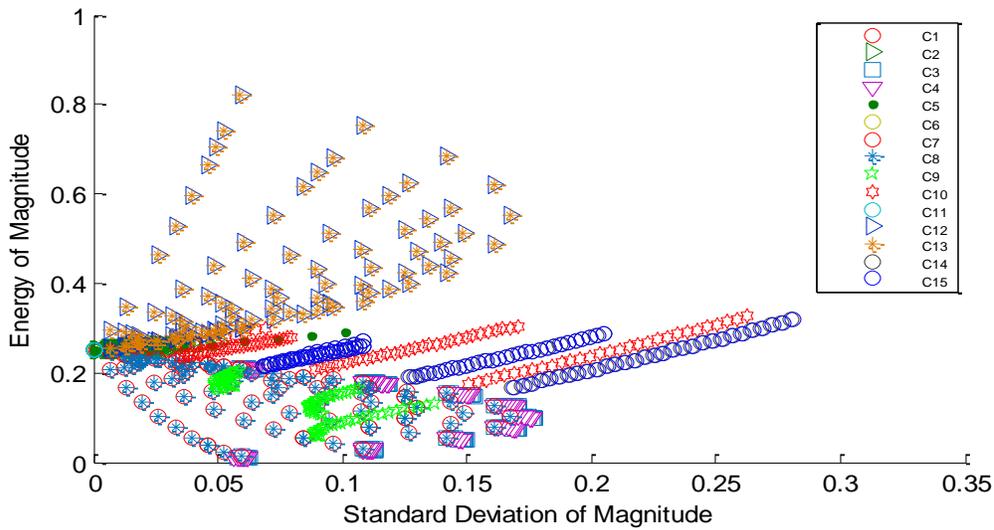

Figure 4 : Feature vector plot between Energy and Standard deviation of Magnitude contour

After the division of dataset into training, validation and testing data , WOA algorithm is started which generates search agents to find optimal values with the search space. After finding the values the trained SVM model is validated on the validation dataset. Figure 4 is the feature vector plot of energy and standard deviation of magnitude vector for all 15 types of power quality disturbance chosen in this paper.

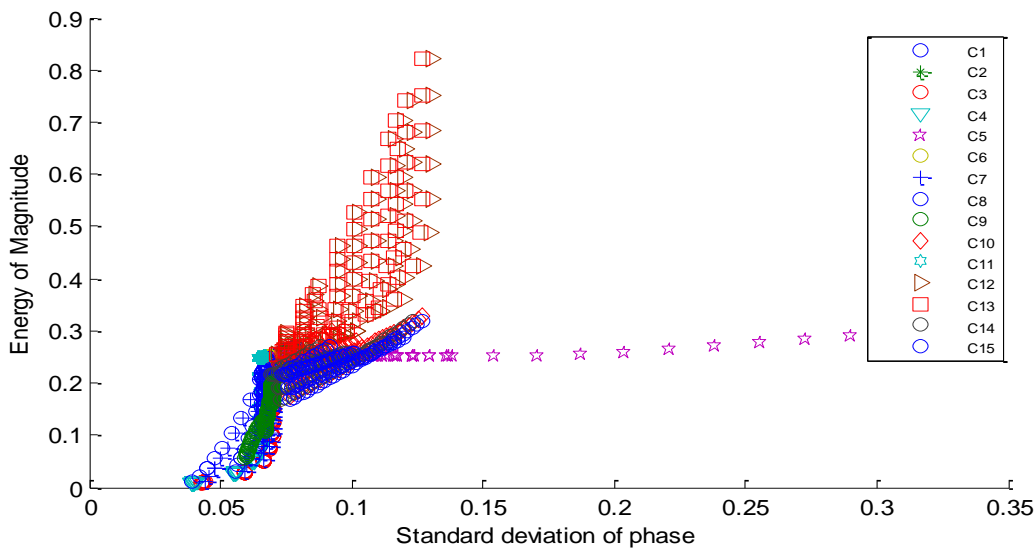

Figure 5 : Feature vector plot between Energy of Magnitude and Standard deviation of Phase contour

**Table 1 : Mathematical Model of different Power Quality Disturbances**

| Class | PQ Event | Numerical Model | Parameters |
|---|---|---|---|
| C1 | Pure Sine | $V(t) = A \sin(\omega t)$ | $A = 1(pu), \omega = 2\pi 50 \, rad/\sec$ |
| C2 | Sag | $V(t) = (1 - \alpha(u(t-t_1) - u(t-t_2))) \sin \omega t$ | $0.1 \leq \alpha \leq 0.9, T \leq t_2 - t_1 \leq 9T$ |
| C3 | Sag with Harmonics | $V(t) = (1 - \alpha(u(t-t_1) - u(t-t_2))) \sin \omega t \times$ $(\alpha_1 \sin(\omega t) + \alpha_3 \sin(3\omega t) + \alpha_5 \sin(5\omega t) + \alpha_7 \sin(7\omega t))$ | $0.1 \leq \alpha \leq 0.9, T \leq t_2 - t_1 \leq 9T$ $0.05 \leq \alpha_3, \alpha_5, \alpha_7 \leq 0.15, \sum \alpha_i^2 = 1$ |
| C4 | Swell | $V(t) = (1 + \alpha(u(t-t_1) - u(t-t_2))) \sin \omega t$ | $0.1 \leq \alpha \leq 0.8, T \leq t_2 - t_1 \leq 9T$ |
| C5 | Swell with Harmonics | $V(t) = (1 + \alpha(u(t-t_1) - u(t-t_2))) \sin \omega t \times$ $(\alpha_1 \sin(\omega t) + \alpha_3 \sin(3\omega t) + \alpha_5 \sin(5\omega t) + \alpha_7 \sin(7\omega t))$ | $0.1 \leq \alpha \leq 0.9, T \leq t_2 - t_1 \leq 9T$ $0.05 \leq \alpha_3, \alpha_5, \alpha_7 \leq 0.15, \sum \alpha_i^2 = 1$ |
| C6 | Interruption | $V(t) = (1 - \alpha(u(t-t_1) - u(t-t_2))) \sin \omega t$ | $0.9 \leq \alpha \leq 1, T \leq t_2 - t_1 \leq 9T$ |
| C7 | Interruption With Harmonics | $V(t) = (1 - \alpha(u(t-t_1) - u(t-t_2))) \sin \omega t \times$ $(\alpha_1 \sin(\omega t) + \alpha_3 \sin(3\omega t) + \alpha_5 \sin(5\omega t) + \alpha_7 \sin(7\omega t))$ | $0.9 \leq \alpha \leq 1, T \leq t_2 - t_1 \leq 9T$ $0.05 \leq \alpha_3, \alpha_5, \alpha_7 \leq 0.15, \sum \alpha_i^2 = 1$ |
| C8 | Oscillatory Transient | $V(t) = \sin \omega t + \alpha e^{\frac{(t-t_1)}{\tau}} \sin \omega_n (t-t_1) \{u(t_2) - u(t_1)\}$ | $0.1 \leq \alpha \leq 0.8, 0.5T \leq t_2 - t_1 \leq 3T$ $8ms \leq \tau \leq 40ms, 300 \leq f_n \leq 900 Hz$ |
| C9 | Flicker | $V(t) = (1 + \alpha_f \sin(\beta \omega t)) \sin \omega t$ | $0.1 \leq \alpha \leq 0.2, 5 \leq \beta \leq 20$ |
| C10 | Flicker with Harmonics | $V(t) = (1 + \alpha_f \sin(\beta \omega t)) \sin \omega t \times$ $(\alpha_1 \sin(\omega t) + \alpha_3 \sin(3\omega t) + \alpha_5 \sin(5\omega t) + \alpha_7 \sin(7\omega t))$ | $0.1 \leq \alpha \leq 0.2, 5 \leq \beta \leq 20$ $0.05 \leq \alpha_3, \alpha_5, \alpha_7 \leq 0.15, \sum \alpha_i^2 = 1$ |
| C11 | Spike | $V(t) = \sin(\omega t) + sign(\sin(\omega t)) \times$ $\left[ \sum_{n=0}^{9} K \times \{u(t-(t_1+0.02n)) - u(t-(t_2+0.02n))\} \right]$ | $0.1 \leq K \leq 0.4, 0 \leq t_1, t_2 \leq 0.5T$ $0.01T \leq t_2, t_1 \leq 0.05T$ |
| C12 | Swell with Interruption | $V(t) = (1 + \alpha_1(u(t-t_1) - u(t-t_2)) - \alpha_2(u(t-t_3) - u(t-t_4))) \sin \omega t$ | $0.1 \leq \alpha_1 \leq 0.8, T \leq t_2 - t_1 \leq 4T$ $0.9 \leq \alpha_2 \leq 1, T \leq t_4 - t_2 \leq 4T$ |
| C13 | Swell with Sag | $V(t) = (1 + \alpha_1(u(t-t_1) - u(t-t_2)) - \alpha_2(u(t-t_3) - u(t-t_4))) \sin \omega t$ | $0.1 \leq \alpha_1 \leq 0.8, T \leq t_2 - t_1 \leq 4T$ $0.1 \leq \alpha_2 \leq 0.8, T \leq t_4 - t_2 \leq 4T$ |
| C14 | Sag with Interruption | $V(t) = (1 - \alpha_1(u(t-t_1) - u(t-t_2)) - \alpha_2(u(t-t_3) - u(t-t_4))) \sin \omega t$ | $0.1 \leq \alpha_1 \leq 0.8, T \leq t_2 - t_1 \leq 4T$ $0.9 \leq \alpha_2 \leq 1, T \leq t_4 - t_2 \leq 4T$ |
| C15 | Sag with harmonic with Interruption | $((1 - \alpha_2(u(t-t_1) - u(t-t_2)) - \alpha_4(u(t-t_3) - u(t-t_4))) \sin \omega t) \times$ $(\alpha_1 \sin(\omega t) + \alpha_3 \sin(3\omega t) + \alpha_5 \sin(5\omega t) + \alpha_7 \sin(7\omega t))$ | $0.1 \leq \alpha_1 \leq 0.8, T \leq t_2 - t_1 \leq 4T$ $0.9 \leq \alpha_2 \leq 1, T \leq t_4 - t_2 \leq 4T$ $0.05 \leq \alpha_3, \alpha_5, \alpha_7 \leq 0.15, \sum \alpha_i^2 = 1$ |

For proper training of classifiers, it is necessary that appropriate signals are generated which should cover most practical scenarios. The mathematical equations given in table 1 is utilized to generate the power quality disturbance

## 6.1 Result Analysis

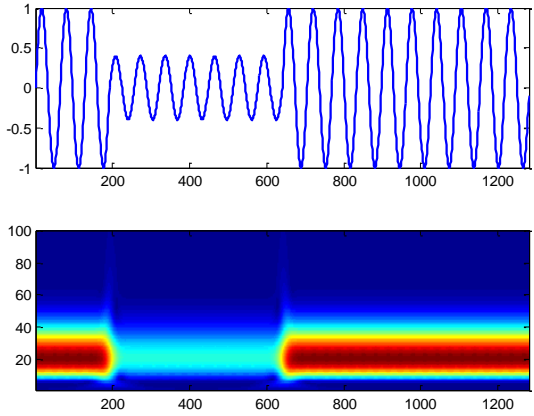

Figure 6 : Sag and corresponding SOGW-ST contour

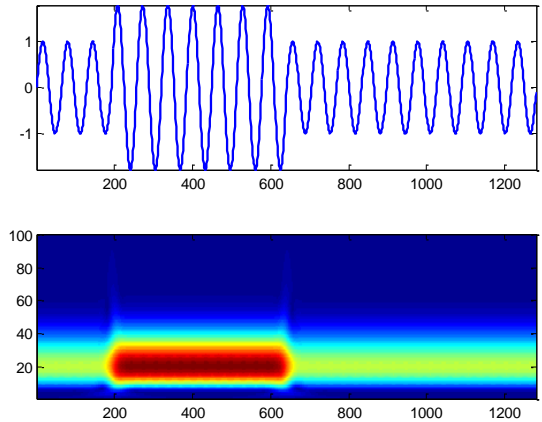

Figure 7 : Swell and corresponding SOGW-ST contour

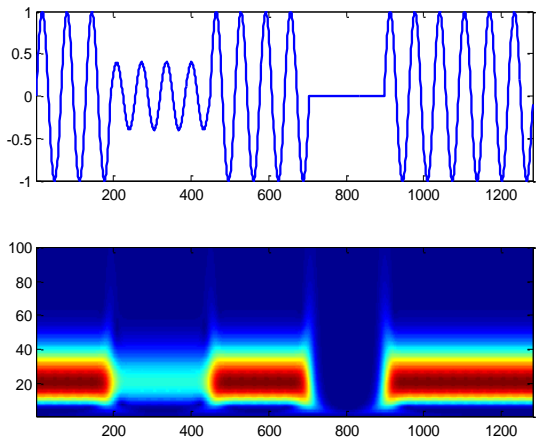

Figure 8 : Sag + Interrupt and SOGW-ST contour

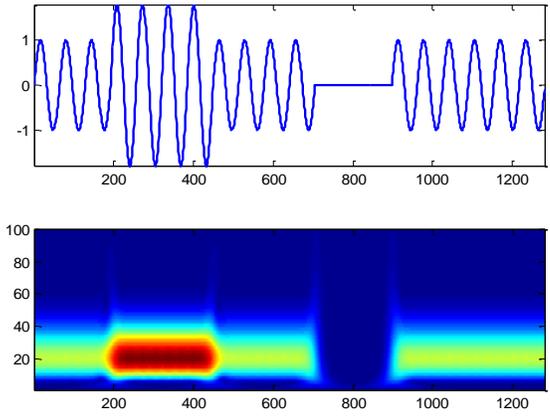

Figure 9: Swell + Interrupt and SOGW-ST contour

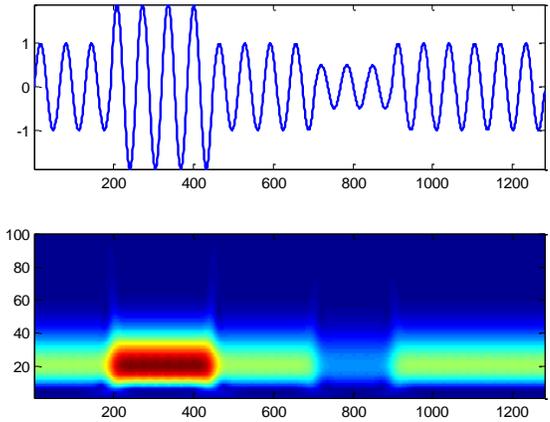

Figure 10 : Swell with sag and SOGW-ST contour

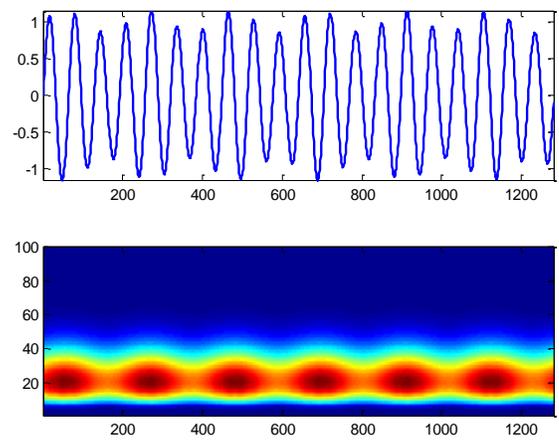

Figure 11 : Flicker and SOGW-ST contour

In the figures given in section 6, it can be clearly seen that whenever there is a change in the structure of the signal a similar change is seen in the ST contour. In Figure 6, at start of sag the ST contour shows a sharp deviation then during the entire sag cycle it squeezes and as soon as the sag end there is again a sharp change in ST contour finally it returns to its original shape when the signal returns to its normal size. From figure 7 to figure 11, it can be observed that the change in the ST contour is according to the change in the original signal that is for different PQ events the ST contour is different. Thus it can be concluded that S-transform successfully detects the PQ deviations.

Various PQ signals are passed through S-transform and the corresponding ST matrix is obtained from which 4 features are extracted. From the ST matrix, the energy and standard deviation of magnitude and phase contour are extracted as features. After feature extraction from ST matrix, the data is used to train various classifiers.

Table 2: Confusion matrix of SVM under noiseless condition

| PQD | C1 | C2 | C3 | C4 | C5 | C6 | C7 | C8 | C9 | C10 | C11 | C12 | C13 | C14 | C15 |
|---|---|---|---|---|---|---|---|---|---|---|---|---|---|---|---|
| 1 | **40** | 0 | 0 | 0 | 0 | 0 | 0 | 0 | 0 | 0 | 0 | 0 | 0 | 0 | 0 |
| 2 | 1 | **40** | 0 | 0 | 0 | 0 | 0 | 0 | 0 | 0 | 0 | 0 | 0 | 0 | 0 |
| 3 | 0 | 0 | **34** | 0 | 0 | 0 | 0 | 0 | 6 | 0 | 0 | 0 | 0 | 0 | 0 |
| 4 | 0 | 0 | 5 | **35** | 0 | 0 | 0 | 0 | 0 | 0 | 0 | 0 | 0 | 0 | 0 |
| 5 | 0 | 0 | 0 | 0 | **40** | 0 | 0 | 0 | 0 | 0 | 0 | 0 | 0 | 0 | 0 |
| 6 | 0 | 0 | 0 | 0 | 0 | **40** | 0 | 0 | 0 | 0 | 0 | 0 | 0 | 0 | 0 |
| 7 | 0 | 0 | 0 | 0 | 0 | 0 | **40** | 0 | 0 | 0 | 0 | 0 | 0 | 0 | 0 |
| 8 | 0 | 0 | 0 | 0 | 0 | 0 | 0 | **40** | 0 | 0 | 0 | 0 | 0 | 0 | 0 |
| 9 | 0 | 0 | 0 | 0 | 0 | 0 | 0 | 0 | **40** | 0 | 0 | 0 | 0 | 0 | 0 |
| 10 | 0 | 0 | 0 | 0 | 0 | 0 | 0 | 0 | 0 | **40** | 0 | 0 | 0 | 0 | 0 |
| 11 | 0 | 0 | 0 | 0 | 0 | 0 | 0 | 0 | 0 | 0 | **40** | 0 | 0 | 0 | 0 |
| 12 | 0 | 0 | 0 | 0 | 0 | 0 | 0 | 0 | 0 | 0 | 0 | **40** | 0 | 0 | 0 |
| 13 | 0 | 0 | 0 | 0 | 0 | 0 | 0 | 0 | 0 | 0 | 0 | 0 | **40** | 0 | 0 |
| 14 | 0 | 0 | 0 | 0 | 0 | 0 | 0 | 0 | 0 | 0 | 0 | 0 | 0 | **40** | 0 |
| 15 | 0 | 0 | 0 | 0 | 0 | 0 | 0 | 0 | 0 | 0 | 0 | 0 | 0 | 0 | **40** |
| **Overall Accuracy = 98.16%** | | | | | | | | | | | | | | | |

Table 3: Confusion Matrix of GA tuned SVM under noiseless condition

| PQD | C1 | C2 | C3 | C4 | C5 | C6 | C7 | C8 | C9 | C10 | C11 | C12 | C13 | C14 | C15 |
|---|---|---|---|---|---|---|---|---|---|---|---|---|---|---|---|
| 1 | **40** | 0 | 0 | 0 | 0 | 0 | 0 | 0 | 0 | 0 | 0 | 0 | 0 | 0 | 0 |
| 2 | 0 | **40** | 0 | 0 | 0 | 0 | 0 | 0 | 0 | 0 | 0 | 0 | 0 | 0 | 0 |
| 3 | 0 | 0 | **40** | 0 | 0 | 0 | 0 | 0 | 0 | 0 | 0 | 0 | 0 | 0 | 0 |
| 4 | 0 | 0 | 0 | **40** | 0 | 0 | 0 | 0 | 0 | 0 | 0 | 0 | 0 | 0 | 0 |
| 5 | 0 | 0 | 0 | 0 | **36** | 0 | 0 | 0 | 4 | 0 | 0 | 0 | 0 | 0 | 0 |
| 6 | 0 | 0 | 0 | 0 | 0 | **40** | 0 | 0 | 0 | 0 | 0 | 0 | 0 | 0 | 0 |
| 7 | 0 | 0 | 0 | 0 | 0 | 0 | **40** | 0 | 0 | 0 | 0 | 0 | 0 | 0 | 0 |
| 8 | 0 | 0 | 0 | 0 | 0 | 0 | 0 | **40** | 0 | 0 | 0 | 0 | 0 | 0 | 0 |
| 9 | 0 | 0 | 0 | 0 | 0 | 0 | 0 | 0 | **40** | 0 | 0 | 0 | 0 | 0 | 0 |
| 10 | 0 | 0 | 0 | 0 | 0 | 0 | 0 | 0 | 0 | **40** | 0 | 0 | 0 | 0 | 0 |
| 11 | 0 | 0 | 0 | 0 | 0 | 0 | 0 | 0 | 0 | 0 | **40** | 0 | 0 | 0 | 0 |
| 12 | 0 | 0 | 0 | 0 | 0 | 0 | 0 | 0 | 0 | 0 | 0 | **40** | 0 | 0 | 0 |
| 13 | 0 | 0 | 0 | 0 | 0 | 0 | 0 | 0 | 0 | 0 | 0 | 0 | **40** | 0 | 0 |
| 14 | 0 | 0 | 0 | 0 | 0 | 0 | 0 | 0 | 0 | 0 | 0 | 0 | 1 | **36** | 3 |
| 15 | 0 | 0 | 0 | 0 | 0 | 0 | 0 | 0 | 0 | 0 | 0 | 0 | 0 | 0 | **20** |
| **Overall Accuracy = 98.67%** | | | | | | | | | | | | | | | |

While training the SVM classifier manually without the use of any meta heuristic algorithm various value values of 'c' and 'gamma' are utilized with the upper and lower bound set as 0.01 and 200000 which are same values used in the meta heuristic algorithms used to find the optimal value of the hyper parameters. It must be noted that with manual tuning the accuracy did not improve more than 98.16% with many different combination of values with the upper and lower bound. In table 2 the confusion matrix of manually tuned SVM is given. It can be observed from the table that for the event sag with harmonics denoted as C3 from the 40 sag signals 6 are misclassified as flicker denoted by C9 similarly 5 swell signals are misclassified as sag with harmonics.. Overall from 600 test signals, 11 are misclassified thereby the overall accuracy of SVM as 98.16%. In table 3 the confusion matrix of GA tuned SVM is given. From the table it can be observed that 4 swell with harmonics signals are misclassified as flicker and 4 sag with interruption signals are misclassified as 3 swell with sag and 1 swell with harmonics with interruption . Overall from 600 test signals, 6 are misclassified with the overall accuracy of 98.67%.

Table 4: Confusion matrix of PSO tuned SVM under noiseless condition

| PQD | C1 | C2 | C3 | C4 | C5 | C6 | C7 | C8 | C9 | C10 | C11 | C12 | C13 | C14 | C15 |
|---|---|---|---|---|---|---|---|---|---|---|---|---|---|---|---|
| 1 | **40** | 0 | 0 | 0 | 0 | 0 | 0 | 0 | 0 | 0 | 0 | 0 | 0 | 0 | 0 |
| 2 | 0 | **40** | 0 | 0 | 0 | 0 | 0 | 0 | 0 | 0 | 0 | 0 | 0 | 0 | 0 |
| 3 | 0 | 0 | **40** | 0 | 0 | 0 | 0 | 0 | 0 | 0 | 0 | 0 | 0 | 0 | 0 |
| 4 | 0 | 0 | 0 | **40** | 0 | 0 | 0 | 0 | 0 | 0 | 0 | 0 | 0 | 0 | 0 |
| 5 | 0 | 0 | 0 | 0 | **35** | 5 | 0 | 0 | 0 | 0 | 0 | 0 | 0 | 0 | 0 |
| 6 | 0 | 0 | 0 | 0 | 0 | **40** | 0 | 0 | 0 | 0 | 0 | 0 | 0 | 0 | 0 |
| 7 | 0 | 0 | 0 | 0 | 0 | 0 | **40** | 0 | 0 | 0 | 0 | 0 | 0 | 0 | 0 |
| 8 | 0 | 0 | 0 | 0 | 0 | 0 | 0 | **40** | 0 | 0 | 0 | 0 | 0 | 0 | 0 |
| 9 | 0 | 0 | 0 | 0 | 0 | 0 | 0 | 0 | **40** | 0 | 0 | 0 | 0 | 0 | 0 |
| 10 | 0 | 0 | 0 | 0 | 0 | 0 | 0 | 0 | 0 | **40** | 0 | 0 | 0 | 0 | 0 |
| 11 | 0 | 0 | 0 | 0 | 0 | 0 | 0 | 0 | 0 | 0 | **40** | 0 | 0 | 0 | 0 |
| 12 | 0 | 0 | 0 | 0 | 0 | 0 | 0 | 0 | 0 | 0 | 0 | **40** | 0 | 0 | 0 |
| 13 | 0 | 0 | 0 | 0 | 0 | 0 | 0 | 0 | 0 | 0 | 0 | 0 | **40** | 0 | 0 |
| 14 | 0 | 0 | 0 | 0 | 0 | 0 | 0 | 0 | 0 | 0 | 0 | 0 | 0 | **40** | 0 |
| 15 | 0 | 0 | 0 | 0 | 0 | 0 | 0 | 0 | 0 | 0 | 0 | 0 | 0 | 0 | **40** |
| **Overall Accuracy = 99.16%** | | | | | | | | | | | | | | | |

Table 5 : Confusion matrix of WOA tuned SVM under noiseless condition

| PQD | C1 | C2 | C3 | C4 | C5 | C6 | C7 | C8 | C9 | C10 | C11 | C12 | C13 | C14 | C15 |
|---|---|---|---|---|---|---|---|---|---|---|---|---|---|---|---|
| 1 | **40** | 0 | 0 | 0 | 0 | 0 | 0 | 0 | 0 | 0 | 0 | 0 | 0 | 0 | 0 |
| 2 | 0 | **40** | 0 | 0 | 0 | 0 | 0 | 0 | 0 | 0 | 0 | 0 | 0 | 0 | 0 |
| 3 | 0 | 0 | **40** | 0 | 0 | 0 | 0 | 0 | 0 | 0 | 0 | 0 | 0 | 0 | 0 |
| 4 | 0 | 0 | 2 | **38** | 0 | 0 | 0 | 0 | 0 | 0 | 0 | 0 | 0 | 0 | 0 |
| 5 | 0 | 0 | 0 | 0 | **40** | 0 | 0 | 0 | 0 | 0 | 0 | 0 | 0 | 0 | 0 |
| 6 | 0 | 0 | 0 | 0 | 0 | **40** | 0 | 0 | 0 | 0 | 0 | 0 | 0 | 0 | 0 |
| 7 | 0 | 0 | 0 | 0 | 0 | 0 | **40** | 0 | 0 | 0 | 0 | 0 | 0 | 0 | 0 |
| 8 | 0 | 0 | 0 | 0 | 0 | 0 | 0 | **40** | 0 | 0 | 0 | 0 | 0 | 0 | 0 |
| 9 | 0 | 0 | 0 | 0 | 0 | 0 | 0 | 0 | **40** | 0 | 0 | 0 | 0 | 0 | 0 |
| 10 | 0 | 0 | 0 | 0 | 0 | 0 | 0 | 0 | 0 | **40** | 0 | 0 | 0 | 0 | 0 |
| 11 | 0 | 0 | 0 | 0 | 0 | 0 | 0 | 0 | 0 | 0 | **40** | 0 | 0 | 0 | 0 |
| 12 | 0 | 0 | 0 | 0 | 0 | 0 | 0 | 0 | 0 | 0 | 0 | **40** | 0 | 0 | 0 |
| 13 | 0 | 0 | 0 | 0 | 0 | 0 | 0 | 0 | 0 | 0 | 0 | 0 | **40** | 0 | 0 |
| 14 | 0 | 0 | 0 | 0 | 0 | 0 | 0 | 0 | 0 | 0 | 0 | 0 | 0 | **40** | 0 |
| 15 | 0 | 0 | 0 | 0 | 0 | 0 | 0 | 0 | 0 | 0 | 0 | 0 | 0 | 0 | **40** |
| **Overall Accuracy = 99.66%** | | | | | | | | | | | | | | | |

Table 4 denotes the confusion matrix of PSO tuned SVM. From the confusion matrix, it can be observed that out of the 600 test signals 5 signals are misclassified thereby the overall accuracy is 99.16%. Using Whale Optimization Algorithm for tuning the hyper parameters provided the highest accuracy of 99.66% with two signals being misclassified. This clearly demonstrates superior performance in automatic classification of PQ events if WOA is used to tune the parameters of SVM classifier.

Table 6: Classification accuracy of SVM with various meta-heuristic algorithms under different conditions

|  | SVM | GA-SVM | PSO-SVM | WOA-SVM |
|---|---|---|---|---|
| 30 dB Noise | 96.00 | 98.16 | 98.67 | 99.33 |
| Noiseless | 98.16 | 98.67 | 99.16 | 99.67 |
| Overall Accuracy | 97.08 | 98.41 | 98.91 | **99.50** |

From Table 6, it can be concluded that for classification of power quality events, meta-heuristic optimized SVM outperforms manually tuned SVM and WOA tuned SVM outperforms other widely used meta-heuristic algorithms such as PSO and GA.

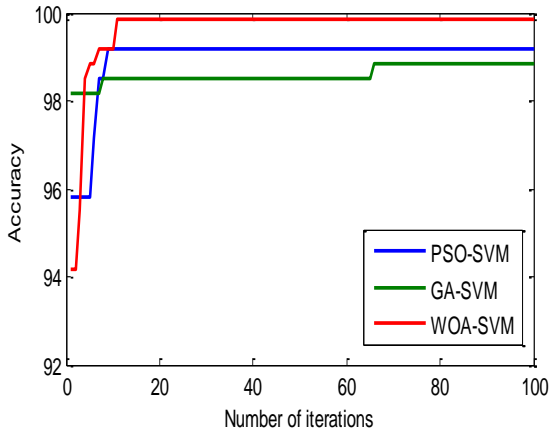 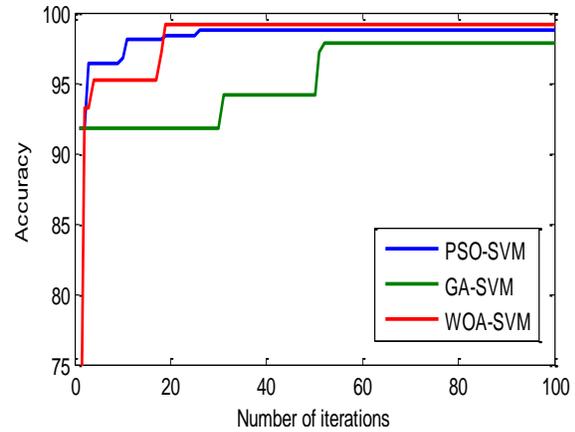

Figure 5: Convergence curve of Classifiers under Noiseless      Figure 6: Convergence curve of Classifiers under 30 dB Noise

Figure 5 and Figure 6 illustrate the convergence curve of meta-heuristic algorithm tuned SVM classifier under noiseless and noisy condition. From the above figures it can be clearly concluded that WOA outperforms other meta-heuristic algorithm.

### 6.2 Circuits for sag, swell and interrupt generation

To further validate the proposed methodology real time signals generated from a simple laboratory setup are used for verifying the detection capability of second order Gaussian window S-Transform. 3 types of Power Quality Disturbances namely Sag, Swell and Interrupt are generated using the laboratory setup. The figure 7 shows the circuit for generating sag and swell similarly figure 8 shows the circuit diagram for generating interrupt.

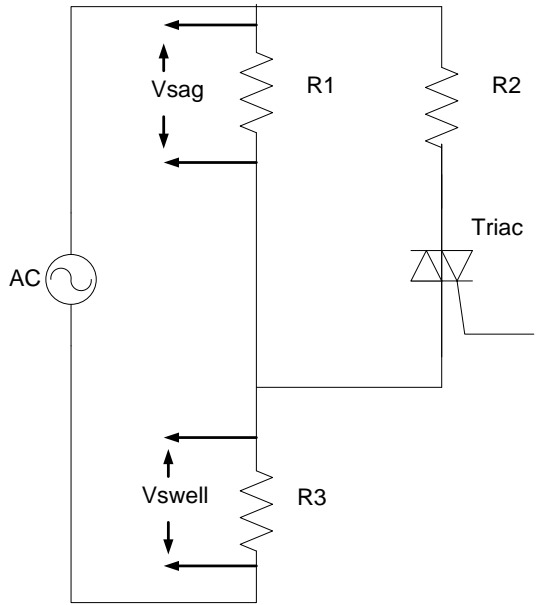 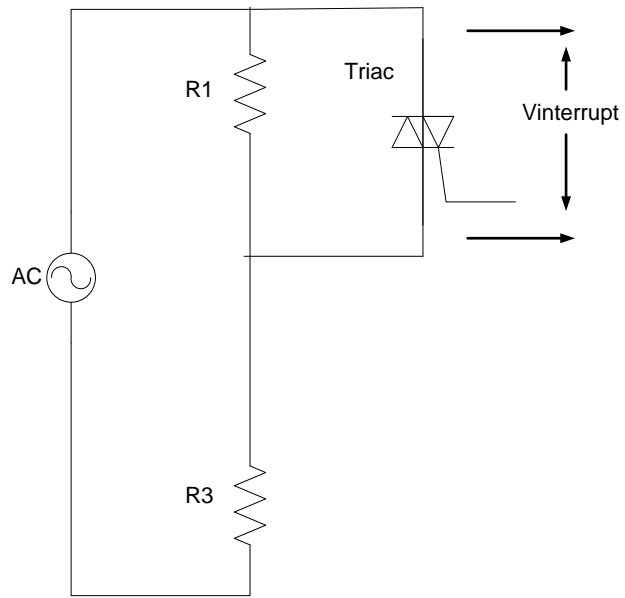

Figure 7: Sag and Swell Generator                Figure 8: Interrupt Generator

In the given circuit diagram above it can be observed that using a simple circuit sag and swell are generated.In figure 7 using appropriate switching gate pulse sag and swell are generated. It must be noted that appropriate values of R1, R2 and R3 are necessary for generation of sag and swell. Once the supply are given and appropriate switching pulse is given  the sag voltage is collected across R1 and the swell voltage is collected across R3. Similarly using the circuit given in figure 8 the interrupt voltage is generated by collecting the voltage across the triac during the application of gate pulse signal.

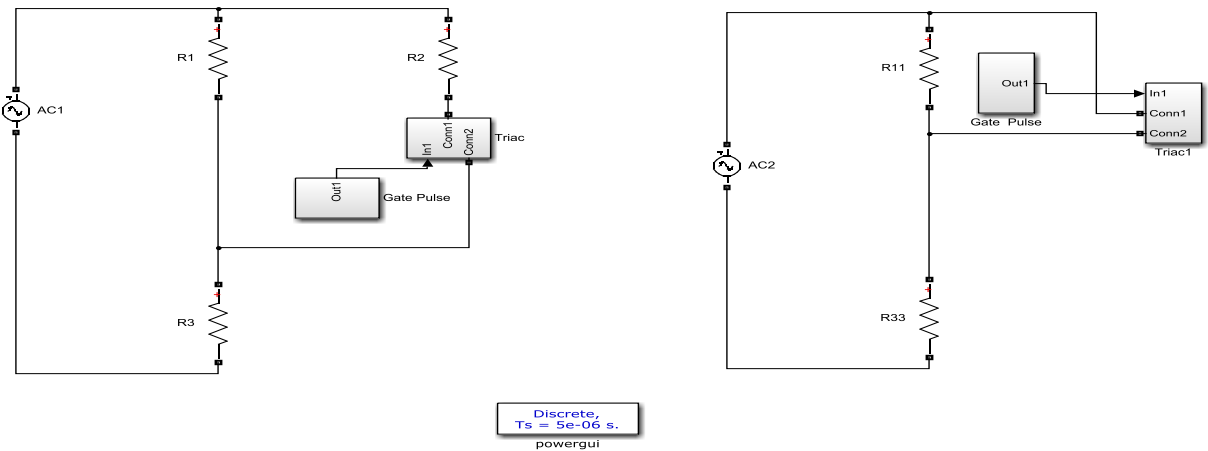

Figure 9: Simulink model of Sag, Swell and interrupt generators

For Simulink implementation of triac we have used two thyristors in anti parallel to each other with the same gate pulse given to both. Since no inherent triac blocks are available in Simulink.

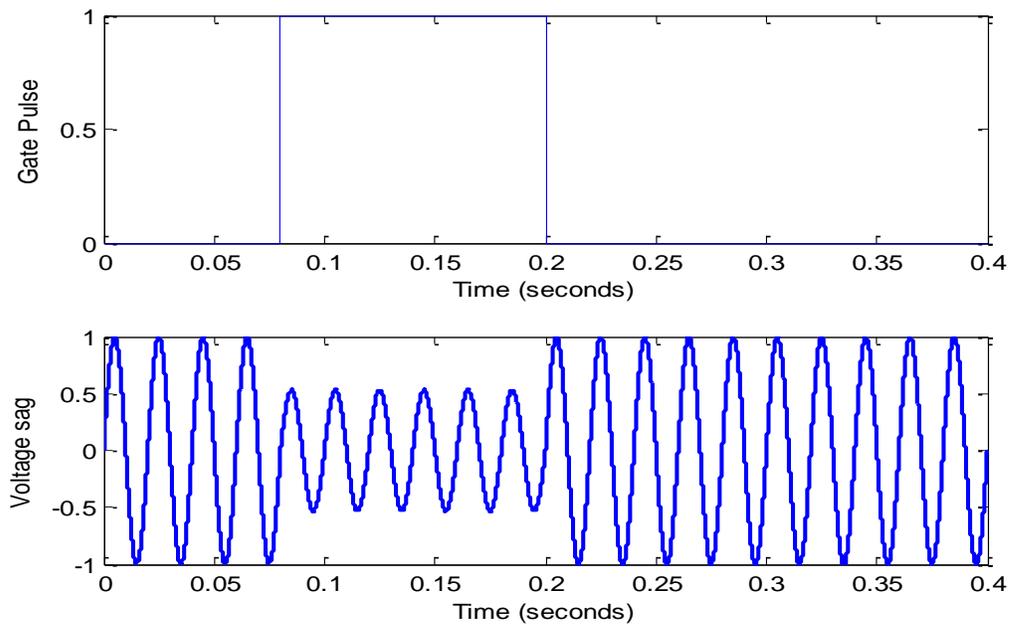

Figure 10: Gate pulse and corresponding sag voltage from Simulink model

From the above figure it can be observed that gate pulse is initiated at an instance between 0.05 and 0.1 and ends at 0.2 sec and at the instance of initiation of gate pulse a voltage sag starts which continues till the end of gate pulse and returns to normal voltage of range of-1 to 1. This illustrates the generation of sag.

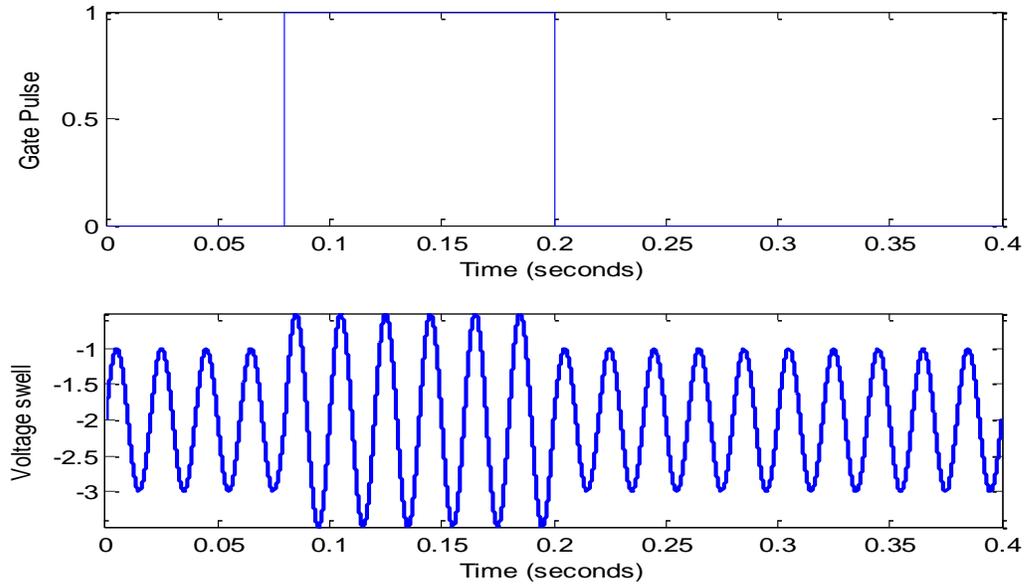

Figure 11: Gate Pulse and the corresponding Swell voltage

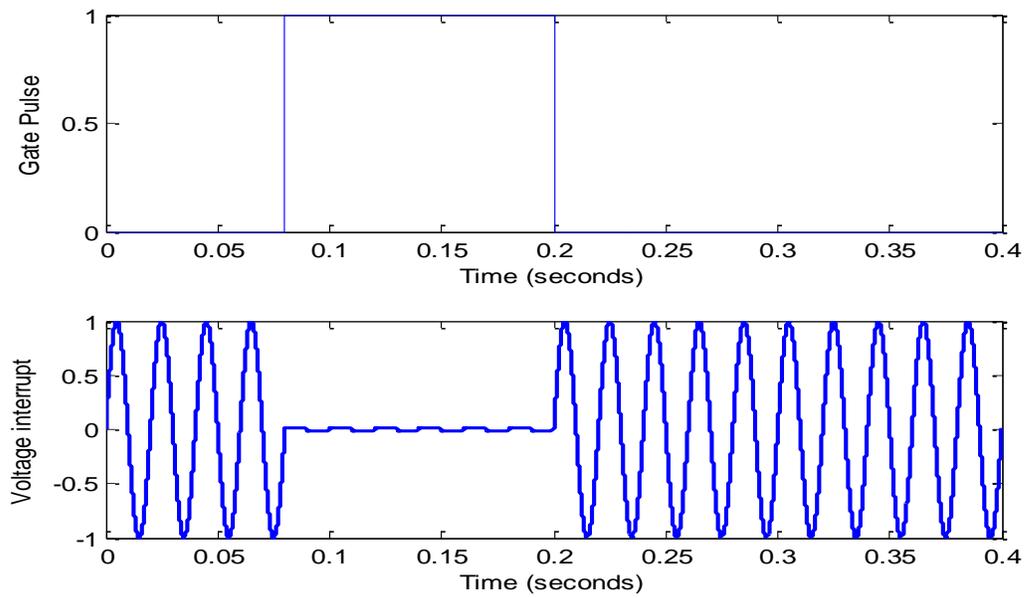

Figure 12: Gate Pulse and the corresponding Interrupt Signal

The figure 11 shows the swell signal which is collected from R3 resistance from shown in figure 7. Similarly the interrupt signal is collected across the triac as given in figure 8. The gate pulse and the corresponding interrupt signal is illustrated in figure 12.

## 6.3 Real time signal generation

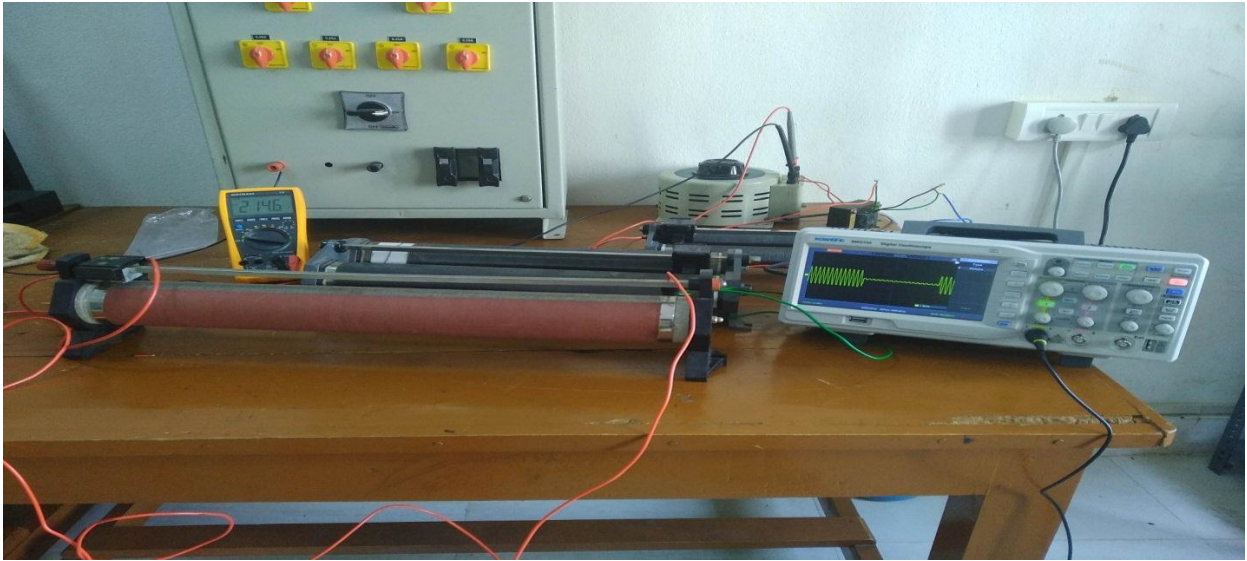

Figure 13 : Experimental setup in the laboratory

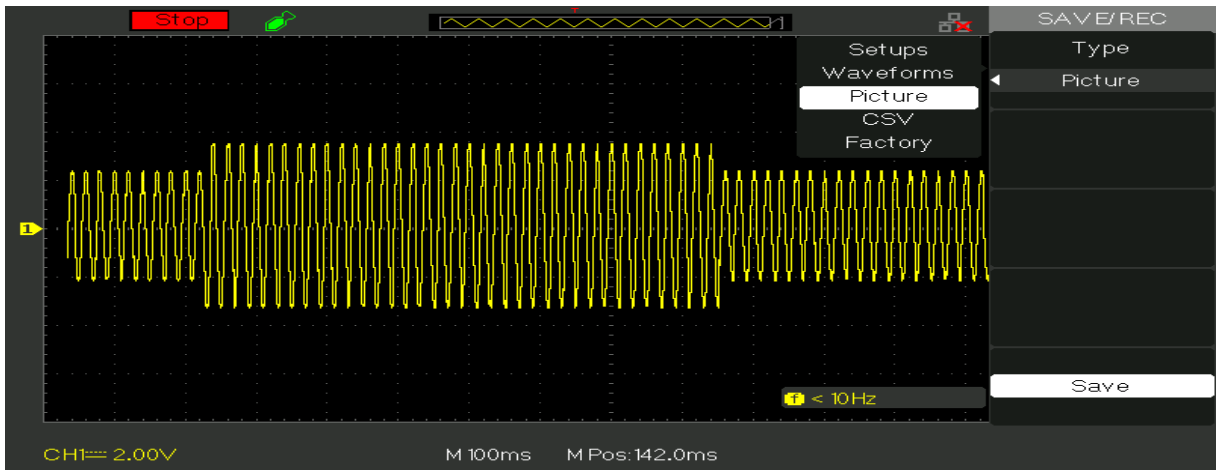

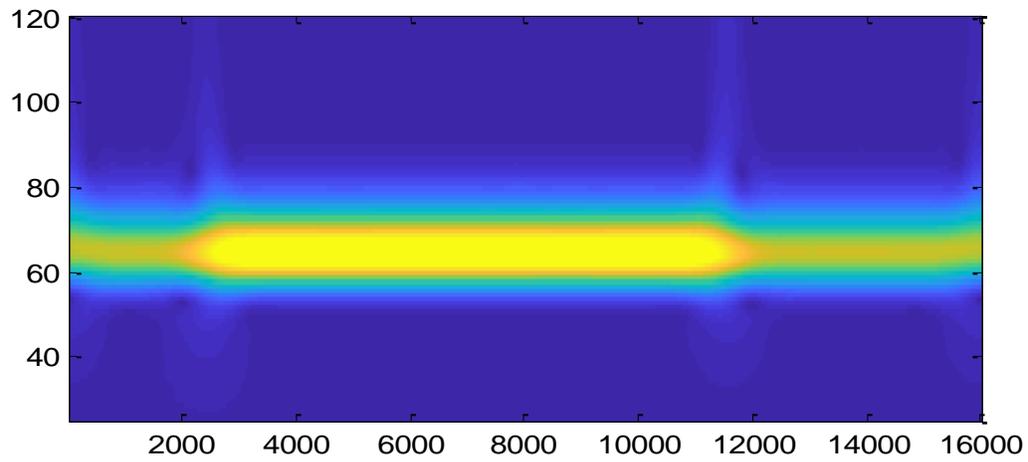

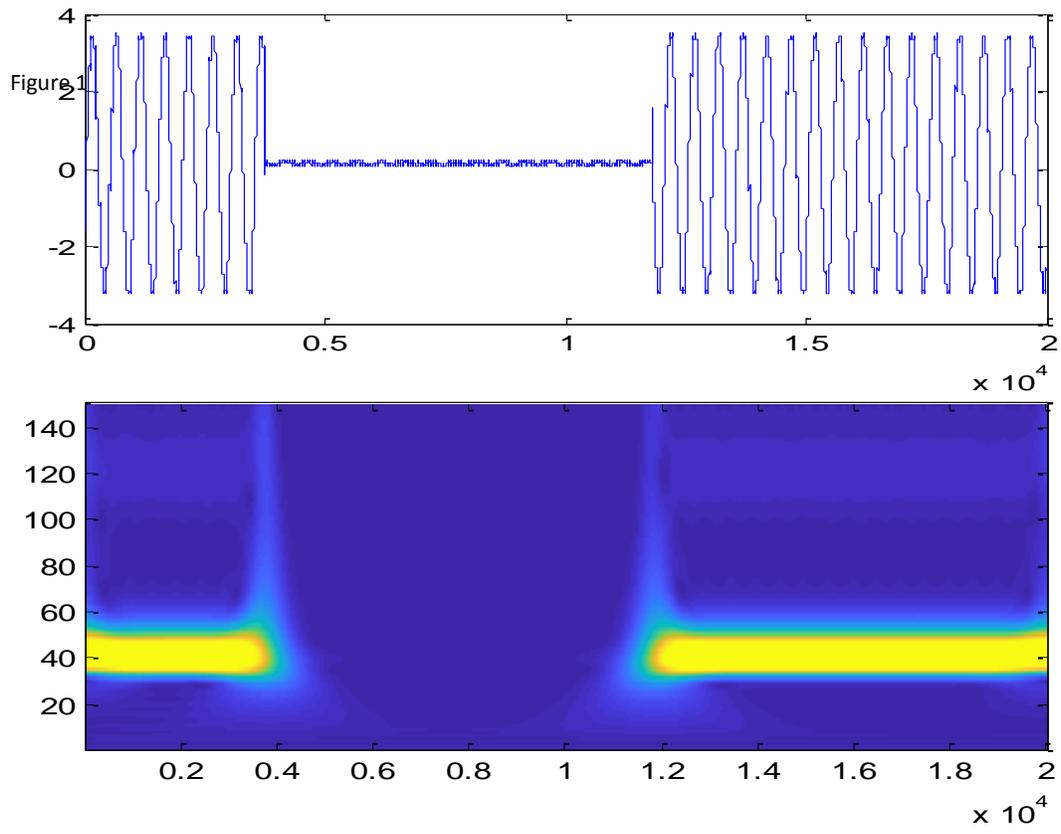

Figure 15: Real time interrupt and the corresponding SOGW-ST contour

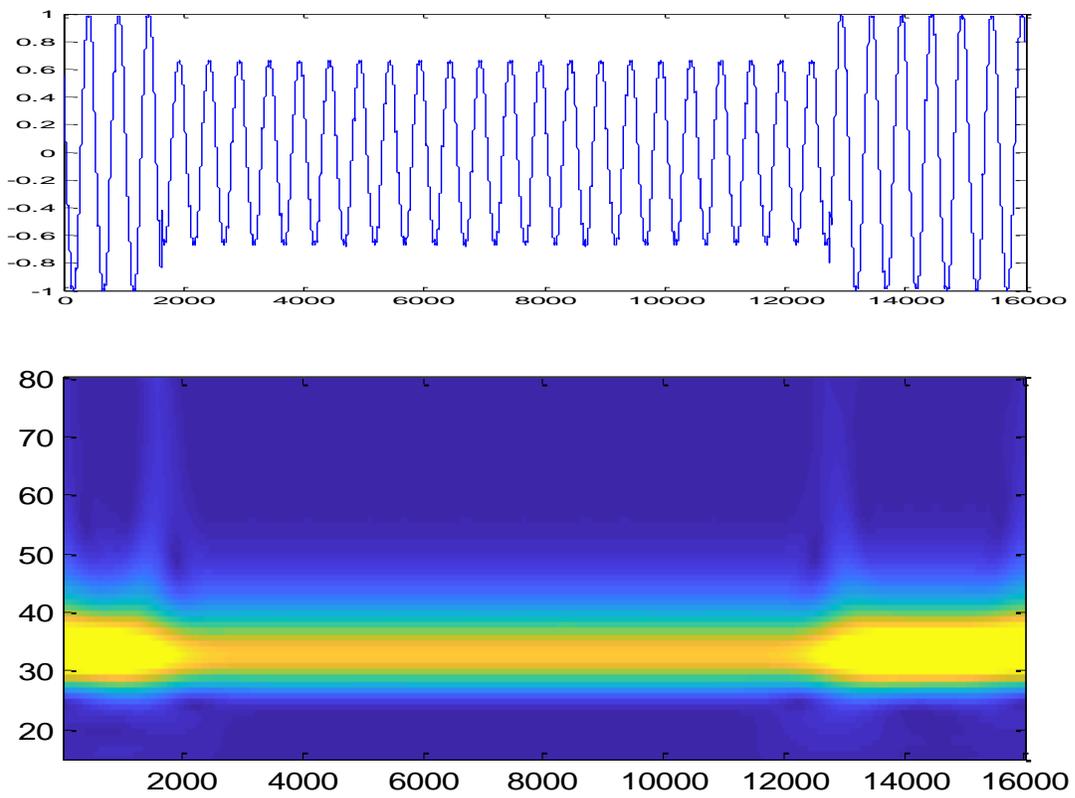

Figure 16: Real time sag and corresponding SOGW-ST contour

Figure 13 illustrates the real time setup for generation of sag, swell and interrupt. In our setup we have utilized 3 rheostats as resistors R1, R2 and R3. Supply is taken from an auto transformer connected to the source. The figure 14 and 15 shows the real time swell and interrupt signals and its corresponding SOGW S-Transform contour. Correspondingly the figure 16 shows the real time sag and the SOGW-ST contour. 10 signals of each type of event is generated. Thus overall 30 signals are taken for testing using a WOA-SVM classifier which was trained using synthetic data. It was observed that its classified 29 signals correctly and 1 sag was misclassified as interrupt with an overall accuracy of 96.67%

Table 7: Comparision of proposed method with other existing methods

| Reference | Feature Extraction Technique | Type of Classifier | No of PQD | Type of Data | Accuracy |
|---|---|---|---|---|---|
| [21] | ST | PNN | 10 | Synthetic | 94.70 |
| [22] | HHT | Balanced Neural Tree | 8 | Synthetic | 97.90 |
| [4] | ST | PNN | 11 | Synthetic | 97.40 |
| [23] | DRST | DAG-SVM | 9 | Synthetic+Noise+Real | 97.77 |
| [5] | ST | DT+NN | 13 | Synthetic | 99.90 |
| [24] | HHT | WBELM | 16 | Synthetic+Real | 97.25 |
| [6] | MOFDST | Random Forest | 13 | Synthetic+Noise+Real | 99.61 |
| Proposed | SOGW-ST | WOA+SVM | 15 | Synthetic+Noise+Real | 99.50 |

The results of the proposed SOGW-ST combined with WOA-SVM have been compared with 7 other methods in existing literature. The table 7 list the various techniques for detection and classification along with the proposed technique. [21] use ST for detection and Probabilistic neural network for classification achieving accuracy of 94.70. Biswal et al. [22] used HHT for detection and a balanced neural tree for classification achieving accuracy of 97.90 %. Kumar et al[5] used decision tree to classify signals into 3 categories using ST matrix and achieved classification accuracy of 99.90% but the efficacy of such a technique has not been evaluated under noisy condition. Reddy et al. [6] maximized the concentration measure under a fixed set of constraints using genetic algorithm and achieved classification accuracy of 99.61% but real time implementation of such a technique can be extremely complex and time consuming. In our proposed method we focus on optimizing the classification accuracy of SVM. It must be noted that once SVM is trained offline and deployed no further need for optimization is necessary. Thus SVM classifier is deployed online with offline evaluated optimal parameters thereby reducing computational complexity.

## 7.Conclusion

In this paper 15 types of Power Quality events are generated for analysis. S-transform is used for feature extraction of PQ events. Then MST contour of various PQ events is developed to clearly illustrate the efficacy of the Modified S-Transform in PQ detection. The corresponding ST matrix is generated from which various features such as energy and standard deviation of the amplitude and phase contour are extracted for the training of machine learning classifiers. From the testing of classifiers, it was observed that WOA tuned SVM provided the highest classification accuracy. WOA tuned SVM is compared with PSO tuned SVM, GA tuned SVM and manually tuned SVM. For real time generation of sag, swell and interrupt two novel circuits are created which are later implemented in lab. The Modified S Transform is applied on real time signals extracted from laboratory and its detection capability is verified. It was observed that the proposed approach provides the highest accuracy on test data verifying that it can be used for automatic PQ classification with superior performance capability.